 \documentclass[final,5p,times,twocolumn]{elsarticle}

\usepackage{amssymb}
\usepackage{graphicx}
\usepackage{dcolumn,color}
\usepackage{bm}
\usepackage{epstopdf}
\usepackage{epsfig}
\usepackage{color,framed}
\usepackage{amsmath,amsthm,amssymb}

\biboptions{comma,square,compress,numbers,sort&compress}
\journal{Physics Letters B}

\begin{document}

\begin{frontmatter}



\title{Non-perturbative procedure for stable $K$-brane}

\author[lzu,IFAE]{Yuan Zhong}
\author[lzu]{Yu-Xiao Liu\corref{cor1}}
  \ead{liuyx@lzu.edu.cn}
  \cortext[cor1]{The corresponding author.}
\author[Qingdao]{Zhen-Hua Zhao}
\address[lzu]{Institute of Theoretical Physics, Lanzhou University,\\
 Lanzhou 730000, People' s Republic of China}
\address[IFAE]{IFAE, Universitat Aut$\grave{o}$noma de Barcelona, 08193 Bellaterra, Barcelona, Spain}
\address[Qingdao]{Department of Applied Physics, Shandong University of Science and Technology,\\
Qingdao 266590, People' s Republic of China}

\begin{abstract}
We propose a novel first-order formalism for a type of $K$-brane systems. An example solution is presented and studied. We illustrate how the noncanonical kinetic term can affect the properties of the model, such as the stability of the solutions, the localization of fermion and graviton. We argue that our solution is stable against linear perturbations. The tensor zero mode of graviton can be localized while the scalar zero mode cannot. The localization condition for fermion is also discussed.
\end{abstract}


\end{frontmatter}


\section{Introduction}
Scalar fields with noncanonical kinetic terms, namely, the $K$-fields were initially introduced to cosmology as a new mechanism of inflation ~\cite{Armendariz-PiconDamourMukhanov1999,GarrigaMukhanov1999,Armendariz-PiconMukhanovSteinhardt2001}.
Since then, $K$-fields are extensively studied in many fields. One of its interesting applications is the modeling of thick $K$-branes. Unlike the standard thick brane models, where branes are usually generated by canonical scalar field(s) (see~\cite{RubakovShaposhnikov1983,RandallSundrum1999,RandallSundrum1999a,
DeWolfeFreedmanGubserKarch2000,Gremm2000,
Gremm2000a,CsakiErlichHollowoodShirman2000,Giovannini2001a,Giovannini2003,
Giovannini2002} for some of the original papers of brane world, and Refs.~\cite{Rubakov2001,Csaki2004,Shifman2010,DzhunushalievFolomeevMinamitsuji2010} for reviews), thick $K$-branes are domain wall branes generated by scalar fields with noncanonical kinetic terms~\cite{BazeiaBritoNascimento2003,KoleyKar2005b,AdamGrandiKlimasSanchez-GuillenWereszczynski2008,BazeiaGomesLosanoMenezes2009,LiuZhongYang2010,CastroMeza2013}.

Like any kind of brane world model, the study of thick $K$-brane models also contains at least three nontrivial issues: solutions, stability, and properties (for example, the localization of bulk matter fields and gravitons). The stability of thick $K$-brane has been generally discussed in Ref.~\cite{ZhongLiu2013}, where the stability conditions for solutions of a large class of thick $K$-brane models were derived.

As to the issue of finding solutions, it is worth to mention that there are interesting dualities between some $K$-field models and the standard model. In other words, some $K$-field models support the same solution given by the model with standard kinetic term of scalar field~\cite{AndrewsLewandowskiTroddenWesley2010}. Such $K$-field models are called the twin-like models of the corresponding standard model, and vice versa. Inspired by the original work~\cite{AndrewsLewandowskiTroddenWesley2010}, some authors studied twin-like models in brane world~\cite{BazeiaDantasGomesLosanoMenezes2011,AdamQueiruga2011,BazeiaLobaoMenezes2012,AdamQueiruga2012}. The twin-like duality offers us an alternative way to find simple analytical $K$-brane solutions. However, some of the models can be distinguished from the standard model only when linear perturbations (especially, the scalar perturbations) are considered, while some are indistinguishable even at linear order~\cite{AdamQueiruga2012}, and hence might be phenomenologically trivial.

In this paper we follow another route to search for analytical solutions, i.e., the first-order formalism. In this formalism, the original second-order Einstein equations are rewritten as some first-order equations of the superpotential (an arbitary function of the background scalar field). The first application of this formalism in thick $K$-brane models was proposed by Bazeia etc.~\cite{BazeiaGomesLosanoMenezes2009} to solve the following two types of models:
\begin{itemize}
  \item type I: $F(X)=X-\alpha X^2$, and
  \item type II: $F(X)=-X^2/2$.
\end{itemize}
Here $X$ and $F(X)$ represent the standard and generalized kinetic terms of the background scalar field, respectively.

Assuming $\alpha$ to be small, the authors of Ref.~\cite{BazeiaGomesLosanoMenezes2009} found some analytical (but not exact) solution for the type I model. The trapping of bulk fermions on the corresponding branes was discussed in Ref.~\cite{CastroMeza2013}.

In the present paper, we report a new first-order formalism that enables us to obtain exact analytical $K$-brane solutions of the type I model. The stability of our solution against liner perturbation as well as the localization of fermion and gravitons are studied. The type II model has been analytically solved in Ref.~\cite{BazeiaGomesLosanoMenezes2009}, so we will omit it here.

In the next section, we briefly review the $K$-brane model and the stability condition for an arbitrary solution. We revisit the type-I model of Ref.~\cite{BazeiaGomesLosanoMenezes2009} in section~\ref{sec3}, where a new first-order formalism is established to solve the system. In particular, we study the Sine-Gordon superpotential as an example, and give the corresponding solution. With this solution, we study how the noncanonical kinetic terms affect the properties of the model, including the localization of fermions (section~\ref{sec4}), and localization of gravitons of both tensor and scalar parts (see section~\ref{sec5}). In the end, we give a brief summary of our results.

\section{A review on $K$-brane and stability conditions}
\label{sec2}
We study the simplest thick $K$-brane model, where a background $K$-field minimally couples with gravity:
\begin{eqnarray}
S=\int d^5 x \sqrt{-g}\left(\frac{1}{2\kappa_5^2}R+\mathcal{L}(\phi,X)\right).
\end{eqnarray}
Here $X\equiv-\frac12g^{MN}\nabla_M\phi\nabla_N\phi$ represents the kinetic term of the background scalar field $\phi$. In the standard model of a thick brane, $\mathcal{L}=X-V(\phi)$, where $V(\phi)$ is an arbitrary potential of the scalar field. The Einstein equations are
\begin{eqnarray}
G_{MN}\equiv R_{MN}- \frac{1}{2}g_{MN}R = \kappa _5^2{T_{MN}},
\end{eqnarray}
where the energy-momentum tensor is defined as\footnote{In this paper, we always use $\mathcal {L}_g$ to denote the derivative of $\mathcal {L}$ with respect to $g$, e.g., $\mathcal {L}_X\equiv \partial \mathcal {L}/\partial X$.}
\begin{eqnarray}
T_{MN}\equiv\frac1{\sqrt{-g}}\frac{\delta S_m}{\delta g^{MN}}={g_{MN}}\mathcal{L} + \mathcal{L}_X\nabla_M\phi \nabla_N\phi.
\end{eqnarray}
In this paper, we always use Latin letters $M, N,\cdots$, as the indices of bulk coordinates, and Greek letters $\mu, \nu, \cdots$, as brane coordinate indices. For simplicity, the extra dimension is labeled as $y\equiv x^5$. Then, the general metric that preserves four-dimensional Poincar\'e symmetry takes the following form:
\begin{eqnarray}
\label{metric}
ds^2 = \textrm{e}^{2A(y)}\eta_{\mu\nu}dx^\mu dx^\nu + dy^2,
\end{eqnarray}
where the four-dimensional Minkowski metric $\eta_{\mu\nu}=\textrm{diag}(-1,+1,+1,+1)$, and $\textrm{e}^{2A(y)}$ is called as the warp factor.
With this metric, we can explicitly write the Einstein equations as
\begin{subequations}
\label{Eqy}
\begin{eqnarray}
\label{eqy1}
-3\partial_y^2 A& =& \kappa _5^2{\mathcal {L}_X}(\partial_y \phi)^2,\\
\label{eqy2}
6(\partial_y A)^2& =& \kappa _5^2({\cal L}+{\mathcal {L}_X}(\partial_y \phi)^2).
\end{eqnarray}
\end{subequations}

The equation of motion for the scalar field is given by
\begin{eqnarray}
(\partial_y ^2\phi)(\mathcal{L}_X+2X \mathcal{L}_{XX})
+\mathcal{L}_\phi-2X\mathcal{L}_{X\phi}=-4\mathcal{L}_X(\partial_y \phi)( \partial_y A).
\end{eqnarray}
This equation can be derived from Eqs.~\eqref{Eqy}. Therefore, only two of the dynamical equations are independent.

In principle, one can find uncountable domain wall solutions with different $\mathcal{L}$ (simply because we cannot fix solutions of $A,~\phi,$ and $V(\phi)$ by using only two independent equations). However, not all the solutions are stable against small perturbations around them. The stability of a general class of $K$-brane models was studied in Ref.~\cite{ZhongLiu2013}, the conclusion is that models with
\begin{eqnarray}
\label{stabCondi}
\mathcal{L}_X>0,\quad \gamma\equiv1+2\frac{\mathcal{L}_{XX} X}{\mathcal{L}_X}>0,
\end{eqnarray}
are always stable against linear perturbations.

\section{The model and first-order formalism}
\label{sec3}
Let us study the following model
\begin{eqnarray}
{\cal L} = X - \alpha {X^2} - V(\phi ),
\end{eqnarray}
where $\alpha$ represents the deviation from the standard model, so let us call it \emph{the deviation parameter}. The deviation parameter can take any value provided that the stability conditions~\eqref{stabCondi} are satisfied. Suppose the scalar field is a kink: $\phi(\pm\infty)=\pm v$ with $v$ a constant, and $\phi(0)=0$. Then, the stability conditions imply a lower bound on the parameter:
\begin{eqnarray}
\label{stab}
\alpha>-\frac1{3k^2 v^2}\equiv \alpha_c.
\end{eqnarray}

The same model was studied in Refs.~\cite{BazeiaLosanoMenezes2008,BazeiaGomesLosanoMenezes2009,AlmeidaBazeiaLosanoMenezes2013}, where the authors assumed that the first-order derivative of the warp factor is an arbitrary function of $\phi$, called the superpotential $W(\phi)$:
\begin{eqnarray}
\partial_y A=-\frac{\kappa_5^2}{3}W(\phi).
\end{eqnarray}
Then the Einstein equations \eqref{Eqy} can be rewritten as
\begin{eqnarray}
&&\partial_y \phi+\alpha(\partial_y\phi)^3=W_\phi,\\
\label{pertFirst}
&&V=\frac12\phi'^2+\frac34\alpha\phi'^4-\frac23\kappa_5^2 W^2.
\end{eqnarray}
The introducing of the superpotential enables one to rewrite the original Einstein equations into some first-order differential equations for $A$ and $\phi$. Therefore, this method is called the first-order formalism.

In the standard model ($\alpha=0$), one can easily find analytical solutions by giving a suitable $W(\phi)$~\cite{BazeiaBritoLosano2006}. For nonvanished $\alpha$, however, the Einstein equations despite remains first-order, are hard to solve analytically.
In Refs.~\cite{BazeiaLosanoMenezes2008,BazeiaGomesLosanoMenezes2009,AlmeidaBazeiaLosanoMenezes2013} the authors solved the system for small deviation parameter $\alpha\ll 1$, and analyzed some properties of the system under this approximation. However, in these papers, numerical calculation is inevitable for large $\alpha$.

Therefore, it is interesting to search for a new approach, from which exact analytical solutions can be obtained even when $\alpha$ is very large.
Our basic assumption is that the noncanonical kinetic term affect only the geometry of the space-time while keep the scalar field unaffected. This statement equivalents to the following assumptions:
\begin{eqnarray}
\label{AW}
\partial_y A=-\frac{\kappa_5^2}{3}(W(\phi)+\alpha Y(\phi)),
\end{eqnarray}
and
\begin{eqnarray}
\label{phiW}
\partial_y \phi&=&{W_\phi }.
\end{eqnarray}
Plugging the above equations into Eq.~\eqref{eqy1} and comparing the coefficients of $\alpha$ and $\beta$, we immediately obtain
\begin{eqnarray}
\label{Zphi}
Y_\phi&=&W_\phi ^3.
\end{eqnarray}
From another Einstein equation~\eqref{eqy2}, we get
\begin{eqnarray}
\label{VW}
V = \frac{1}{2}{W_\phi }^2 + \frac{3}{4}\alpha {W_\phi }^4  - \frac{2}{3}\kappa _5^2{(W + \alpha Y)^2}.
\end{eqnarray}
Therefore, given the form of $W(\phi)$, analytical solutions can be obtained by solving two first-order differential equations~\eqref{AW} and \eqref{phiW} with constraint equation~\eqref{Zphi}. This novel first-order formalism allows us to find some analytical solutions of our model.

To illustrate this, let us consider the Sine-Gordon potential:
\begin{eqnarray}
W=k \phi_0^2\sin\left(\frac{\phi }{\phi_0}\right),
\end{eqnarray}
which leads to a kink-like solution for the scalar field:
\begin{eqnarray}
\phi=\phi_0 \textrm{arcsin}\big(\tanh(k y)\big).
\end{eqnarray}
Without loss of generality, we take the parameter $k$ to be positive, so that $\phi(y\to\pm\infty)=\pm\frac{\pi}{2}\phi_0\equiv \pm v$, and $\alpha_c=-\frac{4}{3\pi^2}\frac1{k^2\phi_0^2}$.

Using the constraint equation~\eqref{Zphi}, we get
\begin{eqnarray}
Y=\frac{1}{12} k^3 \phi _0{}^4 \left(9 \sin \left(\frac{\phi }{\phi _0}\right)+\sin \left(\frac{3 \phi }{\phi _0}\right)\right).
\end{eqnarray}
Then the scalar potential can be obtained:
\begin{eqnarray}
V&=&\frac{1}{2} k^2 \phi_0^2 \cos\left(\frac{\phi }{\phi_0}\right)^2+\frac{3}{4} k^4 \alpha  \phi_0^4 \cos\left(\frac{\phi }{\phi_0}\right)^4\nonumber\\
&-&\frac{k^2 \phi_0^2}{18}\left(6+5 k^2 \alpha  \phi_0^2+k^2 \alpha  \phi_0^2 \cos\left(\frac{2 \phi }{\phi_0}\right)\right)^2 \sin\left(\frac{\phi }{\phi_0}\right)^2.
\end{eqnarray}
For convenience, we have taken the dimensionless quantity $\phi _0^2\kappa ^2 _5=3$. Solving Eq.~\eqref{AW}, we obtain the expression of the warp factor:
\begin{eqnarray}
A&=&-\left(1+\frac{2}{3} k^2 \alpha  \phi _0^2 \right)\ln (\cosh (k y))\nonumber\\
&-&\frac{1}{6} k^2 \alpha  \phi _0^2+\frac{1}{6} k^2 \alpha  \phi _0^2 \textrm{sech}^2(k y).
\end{eqnarray}
The asymptotic behavior of $A$ in the boundary of the extra dimension is
\begin{eqnarray}
\label{asympAlpha}
\lim_{y\to\infty} A= -\left(1+\frac{2}{3} k^2 \alpha  \phi _0^2 \right)k|y|.
\end{eqnarray}
Obviously, the geometry of the bulk space-time is asymptotically anti-de Sitter.

\begin{figure}
\begin{center}
\includegraphics[width=0.45\textwidth]{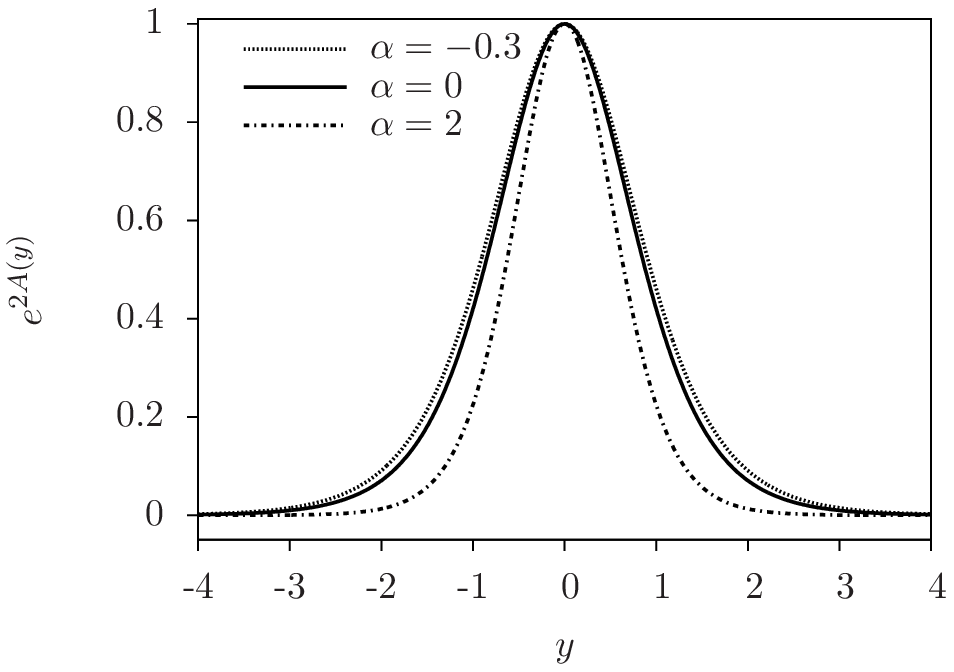}
\includegraphics[width=0.45\textwidth]{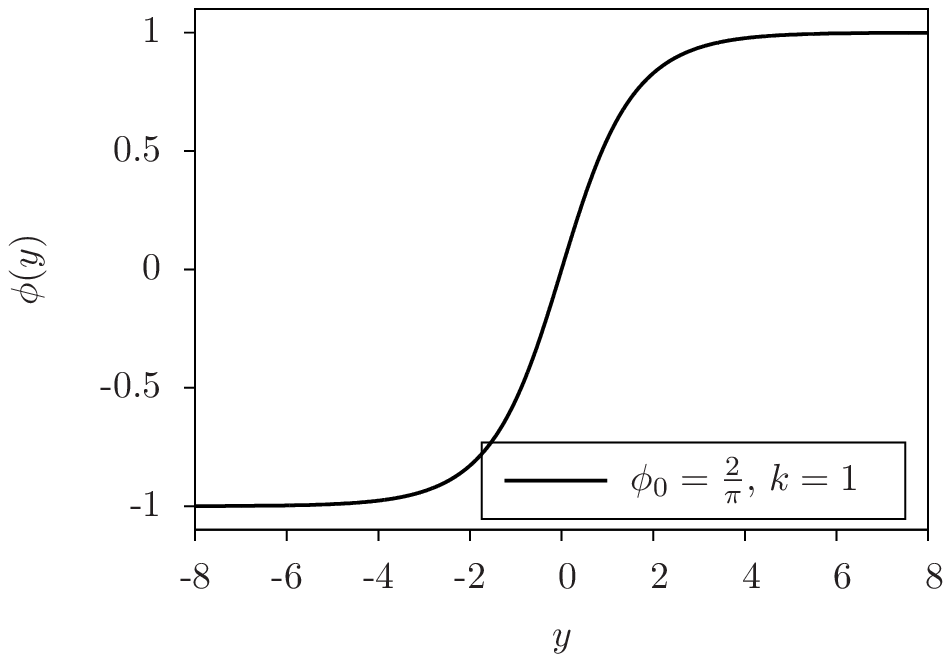}
\includegraphics[width=0.45\textwidth]{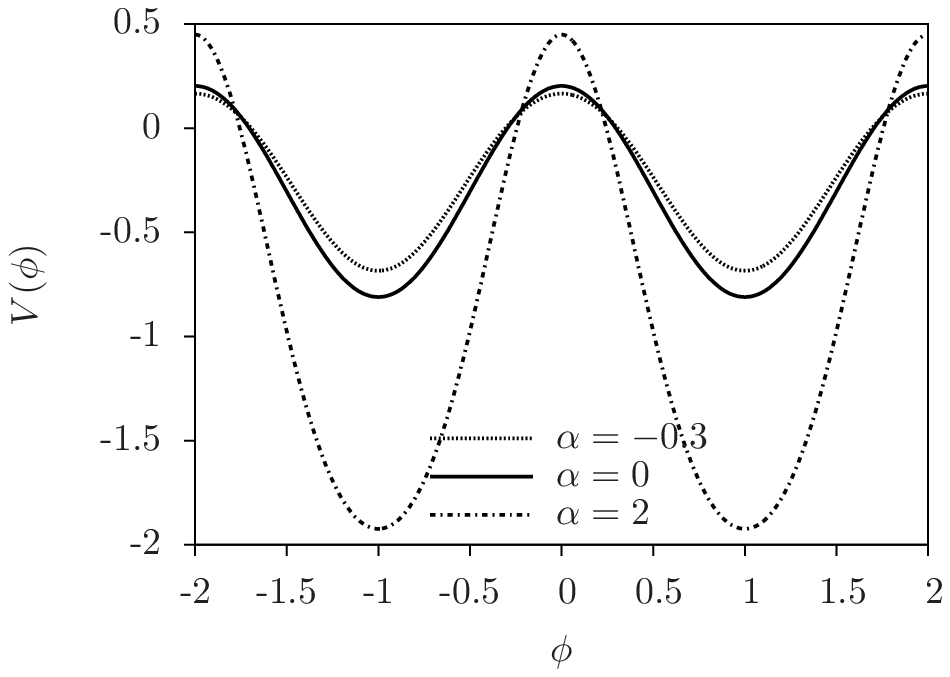}
\includegraphics[width=0.45\textwidth]{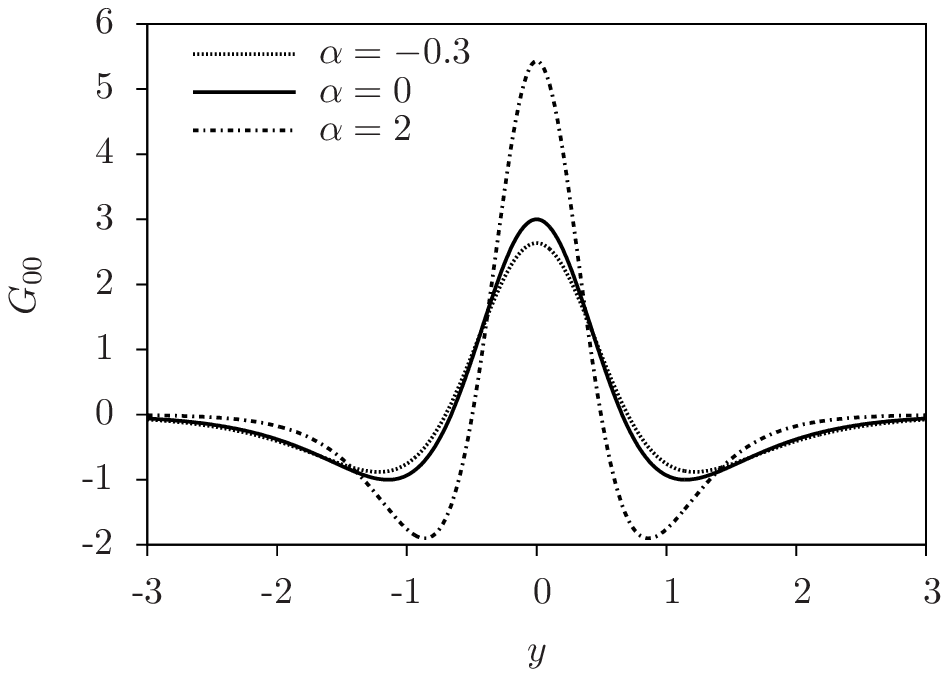}
\end{center}
\caption{Plots of the solutions and $G_{00}$ with $k=1$ and $\phi_0=\frac{2}{\pi}$.} \label{figureProperties}
\end{figure}

Solutions with different values of $\alpha$ are depicted and compared in Fig.~\ref{figureProperties}. The other two parameters are fixed as $k=1$ and $\phi_0=\frac{2}{\pi}$, so that solution with $\alpha>\alpha_c=-1/3$ is stable. Note that instead of studying $T_{00}=-\textrm{e}^{2A}\mathcal{L}$, we prefer to sduty the zero-zero component of the Einstein tensor $G_{MN}$:
\begin{eqnarray}
G_{00}=-3 \textrm{e}^{2 A}\left[2 \left(\partial _yA\right)^2+\partial _y^2 A\right],
\end{eqnarray}
which equivalents to $T_{00}$ (up to a constant), but only depends on the warp factor $A$, so is much easier to calculate.

In the subsequent investigations, it is more convenient to redefine the fifth coordinate as $dy\equiv \textrm{e}^A dr$, and to rewrite line element \eqref{metric} in a conformal flat form
\begin{eqnarray}
\label{metricConformal}
ds^2 = \textrm{e}^{2A(r)}(\eta_{\mu\nu}dx^\mu dx^\nu + dr^2).
\end{eqnarray}
Let us denote the derivative with respect to $r$ by a prime, for example, $A'\equiv\partial_r A$.

\section{Fermion localization and the $\alpha$-term}
\label{sec4}
The issue of trapping fermions with nonvanished $\alpha$ was discussed in Ref.~\cite{CastroMeza2013}, where the analytical background solution is valid only for $0\leq\alpha\ll 1$, and numerical method was applied when $\alpha $ becomes larger. The conclusion of Ref.~\cite{CastroMeza2013} is that the ability
to trap fermions is \emph{inversely proportional} to $\alpha $.
The numerical study for a large range of values of $\alpha $ is also consistent with this conclusion. In this section, we are going to study precisely how a large $\alpha X^2$ term affects on the localization of fermions by using the solution given in section~\ref{sec3}.

As usual, we consider a bulk spin-$\frac12$ field $\Theta(x^\mu,r)$, which couples with gravity and the background scalar in the following form:
\begin{eqnarray}
S_{1/2}=\int d^5x\sqrt{-g}\bar{\Theta}(\Gamma^M D_M-\eta \phi)\Theta.
\end{eqnarray}
Here, $\Gamma^M=(\textrm{e}^{-A}\gamma^\mu,~\textrm{e}^{-A}\gamma^5)$ and $D_M=\partial_M+\omega_M$ are the $\Gamma$-matrixes and covariant derivative in the five-dimensional curved space-time, respectively. $\omega_M=(\frac12A'\gamma_\mu\gamma_5,~0)$ is the spin connection (see Ref.~\cite{LiuZhangZhangDuan2008} for details), and $\eta$ the Yukawa coupling. The equation of motion takes the following form:
\begin{eqnarray}
\label{EOM}
\{\gamma^\mu\partial_\mu+\gamma^5(\partial_r+2A')-\eta \textrm{e}^A\phi\}\Theta=0.
\end{eqnarray}
To obtain the four-dimensional effective action, one needs to decompose the bulk spinor field $\Theta$ into chiral Kaluza-Klein (KK) modes:
\begin{eqnarray}
\label{theta}
\Theta=\textrm{e}^{-2A}\sum_C\sum_n\psi_{C,n}(x^\mu)f_{C,n}(r),
\end{eqnarray}
where $n$ denotes different excitations of KK modes, while $C\in \{+,-\}$ reminds us that each excitation corresponds to two different charities. We assume $\psi_{C,n}=C\gamma^5\psi_{C,n}$, namely, $\psi_{+,n}$ and $\psi_{-,n}$ represent the right- and left-chiral spinor KK modes, respectively, and they are mutually related by four-dimensional Dirac equations:
\begin{eqnarray}
\gamma^\mu\partial_\mu\psi_{C,n}(x^\rho)=m_n\psi_{-C,n}(x^\rho).
\end{eqnarray}

Inserting Eq.~\eqref{theta} into the equation of motion \eqref{EOM}, we obtain a Schr\"odinger-like equation for $f_{C,n}(r)$:
\begin{eqnarray}
(-\partial_r^2+V_C(r))f_{C,n}=m_n^2f_{C,n},
\end{eqnarray}
the potential is
\begin{eqnarray}
V_C&=&(\eta \textrm{e}^A \phi)^2+C\partial_r(\eta \textrm{e}^A \phi).
\end{eqnarray}
Defining $\mathcal{F}\equiv \partial_r+C\eta \textrm{e}^A \phi$, we can rewrite the Schr\"odinger-like equation as follows:
\begin{eqnarray}
\label{decompFermi}
\mathcal{F}\mathcal{F}^\dagger f_{C,n}=m^2_n f_{C,n}.
\end{eqnarray}
According to the supersymmetric quantum mechanics, the above equation ensures that $m^2_n\geq0$. In this paper, what we care is the zero mode $f_{C,0}$ which corresponds to massless fermion ($m_0^2=0$) in four-dimensional space-time. The mass of $f_{C,0}$ is assumed to be generated by the spontaneous symmetry breaking, for example, or by some other mechanisms.

The zero mode can be easily read out from Eq.~\eqref{decompFermi}:
\begin{eqnarray}
f_{C,0}(r)\propto \exp\bigg(C\eta\int_0^r d\bar{r}\textrm{e}^{A(\bar{r})}\phi(\bar{r})\bigg).
\end{eqnarray}
To trap the zero mode on the brane, we require $f_{C,0}$ is normalizable, namely, the integration $\int dr (f_{C,0})^2$ is finite, or
\begin{eqnarray}
\int dy \exp\left(-A(y)+2C\eta\int ^y_0 d\bar{y}\phi(\bar{y})\right)<\infty,
\end{eqnarray}
as written in $y$-coordinate~\cite{LiuZhangZhangDuan2008}. According to Eq.~\eqref{asympAlpha}, the integrand asymptotically behaves as
\begin{eqnarray}
\left(k+\frac23\alpha k^3\phi_0^2+ C\eta\pi\phi_0\right)|y|,\quad \textrm{for} \quad |y|\to +\infty.
\end{eqnarray}
Obviously, the integral converges only when
\begin{eqnarray}
\label{locali}
k+\frac23\alpha k^3\phi_0^2+ C\eta\pi\phi_0<0.
\end{eqnarray}
When $\eta=0$, this condition can be fulfilled by asking $\alpha<-3/(2k^2\phi_0^2)$. However, his violates the stability condition~\eqref{stab}.
For $\eta>0$ and $C=+$, the inequality \eqref{locali} is always violated, so the right-chiral fermion is non-normalizable for positive $\eta$. On the other hand, the left-chiral fermion ($C=-$) can be normalized if
\begin{eqnarray}
\label{localiInequa}
\eta>\frac k{\pi\phi_0}+\frac2{3\pi}\alpha k^3\phi_0>\frac{7k}{9\pi\phi_0}>0.
\end{eqnarray}
To obtain the second inequality, we used Eq.~\eqref{stab}.

In sum, even smaller $\alpha$ can strengthen the localization of fermion, a positive Yukawa coupling is necessary to localize the left-chiral fermion zero mode. Given a fixed coupling $\eta$, a large positive $\alpha$ would destroy the localization condition \eqref{localiInequa}.  Our results are consistent with those of Ref.~\cite{CastroMeza2013}.
\section{Gravitons and the $\alpha$-term}
\label{sec5}
The localization of gravitational modes is another important issue, because it relates to the reproduce and modifications of the four-dimensional Newtonian gravity. To study the localization of gravitational modes, one needs to analyze the spectrum and configurations of small perturbations $\{\delta g_{MN},~\delta\phi\}$ around the background solution $\{g_{MN},~\phi\}$. In $r$-coordinate, we define the perturbed metric as follows:
 \begin{eqnarray}
ds^2 = \textrm{e}^{2A(r)}(\eta_{MN}+h_{MN})dx^M dx^N,
\end{eqnarray}
namely, $\delta g_{MN}\equiv \textrm{e}^{2A(r)}h_{MN}(x^\rho, r)$.

To simplify the calculation, the scalar-tensor-vector decomposition is widely applied in the study of linearization of gravitational systems:
\begin{subequations}
\label{decomposition}
\begin{eqnarray}
{h_{\mu r}} &=& {\partial _\mu }F + {G_\mu },\\
{h_{\mu \nu }} &=& {\eta _{\mu \nu }}\Psi + {\partial _\mu }{\partial _\nu }B + 2{\partial _{(\mu }}{C_{\nu )}} + {D_{\mu \nu }},
\end{eqnarray}
\end{subequations}
where $C_\mu$ and $G_\mu$ are transverse vector perturbations:
\begin{eqnarray}
\partial^\mu C_\mu=0=\partial^\mu G_\mu,
\end{eqnarray}
and $D_{\mu \nu }$ is transverse and traceless (TT) perturbation:
\begin{eqnarray}
\partial^\nu D_{\mu \nu }=0=D^\mu_\mu.
\end{eqnarray}
Note that all indices are raised with $\eta^{\mu\nu}$, so that $\partial^{\mu}\equiv\eta^{\mu\nu}\partial_{\nu}$.

Under this decomposition, the original field perturbations can be classified into scalar ($\Xi\equiv h_{rr},~\Psi,~B$, $F$, and $\Phi\equiv\delta\phi$), tensor ($D_{\mu\nu}$) and vector ($C_\mu$ and $G_\mu$) modes. All these modes are functions of the bulk coordinates $x^\rho$ and $r$. Each type of mode evolves independently~\cite{ZhongLiu2013}, so we can discuss them separately. As in the standard model, the spectrum of the vector modes contains only a nonlocalizable zero mode. So, we omit the vector modes and only consider the tensor and scalar modes.
\subsection{Tensor mode}
Let us first focus on the tensor part, for which the perturbed metric reads:
 \begin{eqnarray}
ds^2 = \textrm{e}^{2A(r)}\left[(\eta_{\mu\nu}+D_{\mu\nu})dx^\mu dx^\nu+dr^2\right].
\end{eqnarray}
Note that the tensor mode is independent with the scalar part and in our model we only modify the scalar lagrangian of the standard model, so the dynamical equation for the tensor mode takes the same form as the one in the standard model~\cite{DeWolfeFreedmanGubserKarch2000,ZhongLiu2013}:
\begin{eqnarray}
  \label{tens}
\square ^{(4)}D_{\mu \nu }+ D_{\mu \nu }'' + 3A'D_{\mu \nu }'&=& 0.
\end{eqnarray}
Consider the following decomposition:
\begin{eqnarray}
D_{\mu \nu }(x^\rho,r)=\textrm{e}^{-3/2A}\epsilon_{\mu\nu}( x^\rho)\chi(r),
\end{eqnarray}
where $\epsilon_{\mu\nu}( x^\rho)$ is transverse and traceless $\eta^{\mu\nu}\epsilon_{\mu\nu}=0=\partial^\mu\epsilon_{\mu\nu}$ and satisfies $\square ^{(4)}\epsilon_{\mu\nu}=m^2 \epsilon_{\mu\nu}$. Then, the KK mode $\chi(r)$ satisfies the following Schr\"odinger-like equation
\begin{eqnarray}
 - \chi'' + {U_T}(r)\chi = {m^2}\chi,
\end{eqnarray}
with
\begin{eqnarray}
{U_T}(r) = \frac{9}{4}A'^2 + \frac{3}{2}A''.
\end{eqnarray}
This equation can be factorized as
\begin{eqnarray}
\label{tensorFactor}
\mathcal{J}\mathcal{J}^\dagger\chi={m^2}\chi,
\end{eqnarray}
where
\begin{eqnarray}
\mathcal{J}\equiv \partial _r+\frac32A',\quad
\mathcal{J}^\dagger=-\partial _r+\frac32A'.
\end{eqnarray}
According to the supersymmeric quantum mechanics, such a factorization ensures that $m^2\geq 0$. So, the model is stable against the tensor perturbation. Meanwhile, the zero mode can be easily read out
\begin{eqnarray}
\chi^{(0)}\propto \textrm{e}^{3/2A}.
\end{eqnarray}
The normalization condition for the zero mode is
\begin{eqnarray}
\int dr  \textrm{e}^{3A(r)}=\int dy \textrm{e}^{2A(y)}<\infty.
\end{eqnarray}
This condition is satisfied, if
\begin{eqnarray}
\alpha  > - \frac{2}{3}\frac{1}{{{k^2}\phi _0^2}}\equiv\alpha_t .
\end{eqnarray}
Recall that the stability condition is $\alpha>\alpha_c=-\frac{4}{3\pi^2}\frac1{k^2\phi_0^2}$. Obviously, $\alpha_c>\alpha_t$, so we can conclude that any solution that satisfies the stability condition supports a localizable tensor zero mode. As a result, four-dimensional Newtonian gravity can be reproduced in these models.

In addition to the zero mode, we have a continuum spectrum which causes a small scale correction to the Newtonian potential. According to Refs.~\cite{CsakiErlichHollowoodShirman2000,BazeiaGomesLosano2009}, the correction is determined by the behavior of ${U_T}(r)$ at large $r$. One can easily proof that the asymptotic behavior of ${U_T}(r)$ is independent of the parameter $\alpha$, so we conclude that the correction to the Newtonian potential is $\Delta \mathcal{V}_{\textrm{Newton}}\propto 1/r^3$ no matter what value $\alpha$ takes.

\begin{figure}
\begin{center}
\includegraphics[width=0.45\textwidth]{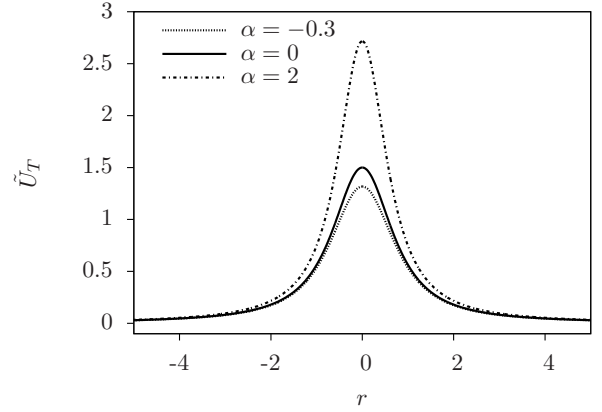}
\end{center}
\caption{Plot of $\tilde{U}_T(r)$ for $k=1$ and $\phi_0=\frac{2}{\pi}$.} \label{figureTensorModePotential}
\end{figure}

The next question is, does the parameter $\alpha$ affects the resonant spectrum of the tensor mode? To illustrate this question, let us study the following equation
\begin{eqnarray}
\label{SUSYtensorFactor}
\mathcal{J}^\dagger\mathcal{J}\tilde{\chi}={m^2}\tilde{\chi}.
\end{eqnarray}
This equation looks like Eq.~\eqref{tensorFactor}, except the order of the operators is reversed. In supersymmetric quantum mechanics, $\tilde{\chi}$ is called as the superpartner of $\chi$. Except the ground state, superpartners share the same spectrum. Thus, if $\tilde{\chi}$ has massive resonant peaks, so does $\chi$. Expanding Eq.~\eqref{SUSYtensorFactor}, we obtain another Schr\"odinger-like equation where the potential is given by
\begin{eqnarray}
\tilde{U}_T(r) = \frac{9}{4}A'^2 - \frac{3}{2}A''.
\end{eqnarray}
The plot of $\tilde{U}_T(r)$ (Fig.~\ref{figureTensorModePotential}) does not show any attractive well, so, it is impossible for $\tilde{\chi}$ to have resonant modes, so dose $\chi$. From the same plot, we also see that just like fermions, gravitons are more likely to be trapped on brane with smaller $\alpha$.

\subsection{Scalar modes}
Let us study the scalar perturbations under the longitude gauge, i.e., take $F=0=B$. This gauge completely fixes the gauge freedoms in the scalar section, and the perturbed metric takes a simple form:
 \begin{eqnarray}
ds^2 = \textrm{e}^{2A(r)}\left[\eta_{\mu\nu}(1+\Psi)dx^\mu dx^\nu+(1+\Xi)dr^2\right].
\end{eqnarray}
Then the perturbation equations are~\cite{ZhongLiu2013}
\begin{eqnarray}
\label{18}
  &&- \Psi - \frac{1}{2}{\Xi} = 0,  \\&&
\label{19}
  \frac{3}{2}A'{\Xi} - \frac{3}{2}\Psi' = \kappa _5^2{\mathcal{L}_{X}}\phi '\Phi,\\&&
\label{20}
\frac{3}{2}{\square ^{(4)}}\Psi - \frac{3}{2}\Psi'' - \frac{3}{2}A'\Psi'
+ \kappa _5^2\phi {'^2}{\mathcal{L}_{XX}}{\textrm{e}^{ - 2A}}\phi {'^2}\Psi \nonumber\\
&&= 2\kappa _5^2{\mathcal{L}_{X}}\phi '\Phi ' - \kappa _5^2\phi {'^2}{\mathcal{L}_{XX}}{\textrm{e}^{ - 2A}}\phi '\Phi ' + \kappa _5^2\phi {'^2}{\mathcal{L}_{X\phi }}\Phi.
\end{eqnarray}
Using Eqs.~\eqref{18}, \eqref{19}, and \eqref{Eqy}, we can eliminate $\Xi$, $\Phi$, and $\mathcal{L}_{X\phi}$ in Eq.~\eqref{20} and obtain the following equation\footnote{Note that the $\partial_y$ in Ref.~\cite{ZhongLiu2013} should be $\partial_r$.}:
\begin{eqnarray}
 && {\square ^{(4)}}\Psi +\gamma\Psi''
  +\gamma\left[\partial _r\ln  \left(\frac{\textrm{e}^{3A}}{\mathcal{L}_X(\phi ')^2}\right)\right]\Psi '\nonumber\\
 & +&2\gamma A' \left[\partial _r\ln  \left(\frac{A'^2}{\mathcal{L}_X(\phi ')^2}\right)\right]\Psi=0,
\end{eqnarray}
where $\gamma=1+2\frac{\mathcal{L}_{XX} X}{\mathcal{L}_X}.$

\begin{figure}
\begin{center}
\includegraphics[width=0.45\textwidth]{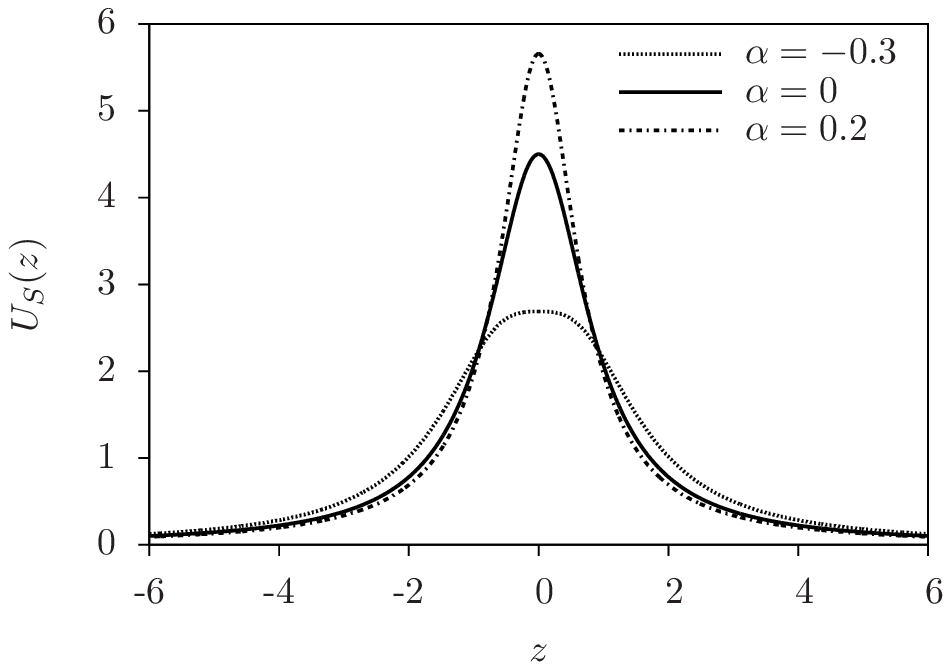}
\end{center}
\caption{Plot of $\tilde{U}_S(z)=\theta\left(\theta ^{-1}\right)^{\centerdot\centerdot}$ for $k=1$ and $\phi_0=\frac{2}{\pi}$.} \label{figureScalarPertPotential}
\end{figure}

In the case $\mathcal{L}_X>0$, we can rewrite the above equation into a more compact form:
\begin{eqnarray}\label{schro}
{\square ^{(4)}}\hat{\Psi}+\gamma\hat{\Psi}''-\gamma \zeta\left(\zeta^{-1}\right)''\hat{\Psi} =0,
\end{eqnarray} by redefining $\Psi=\textrm{e}^{-3A/2}\mathcal{L}_X^{1/2}\phi '\hat{\Psi}$. Here
\begin{equation}
\label{zf}
\zeta\equiv \textrm{e}^{3A/2}\frac{\phi '}{A' }\mathcal{L}_X^{1/2}.
\end{equation}
In addition, if $\gamma>0$, then we can define a new coordinate $z$, such that
\begin{equation}
\label{RWcoord}
\frac{dz}{dr}=\gamma^{-1/2}.
\end{equation}
In the new coordinate, we can rewrite Eq.~\eqref{schro} as
\begin{eqnarray}
{\square ^{(4)}}\hat{\Psi}+\ddot{\hat{\Psi}}-\frac{\dot{\gamma}}{2\gamma}\dot{\hat{\Psi}}
- \zeta\left(\zeta^{-1}\right)^{\centerdot\centerdot}\hat{\Psi}
+\frac{\dot{\gamma}}{2\gamma} \left(\zeta^{-1}\right)^{\centerdot}\zeta\hat{\Psi}=0,
\end{eqnarray}
where dots denote the derivatives with respect to $z$.
After a further redefinition of the field $\hat{\Psi} = \gamma^{1/4}\tilde{\Psi}$, we finally obtain what we are looking for, a Schr\"odinger-like equation:
\begin{eqnarray}
\label{eqPhi}
{\square ^{(4)}}\tilde{\Psi}+\ddot{\tilde{\Psi}}-\tilde{\Psi}\theta \left(\theta ^{-1}\right)^{\centerdot\centerdot}=0,
\end{eqnarray}
where $\theta \equiv\gamma^{1/4}\zeta $. This equation enables us to introduce the KK decomposition
\begin{equation}
\tilde{\Psi}=\sum_n \textrm{e}^{ip_\mu^{n} x^\mu}\varphi_{n}(z).
\end{equation}
Since $(p_\mu^{n})^2=-m_n^2$, Eq.~\eqref{eqPhi} immediately reduces to the following equation for  the  KK mode $\varphi_{n}(z)$:
\begin{eqnarray}
\label{scalarSch}
 - \ddot{\varphi}_{n} + {U_S}(z)\varphi_{n} = m^2_{n}\varphi_{n},
\end{eqnarray}
where ${U_S}(z) = \theta \left(\theta ^{-1}\right)^{\centerdot\centerdot}$.
Equivalently, Eq.~\eqref{scalarSch} can be factorized as follows:
\begin{equation}
\label{Psi}
\mathcal{A}^\dagger\mathcal{A} \varphi_{n}(z)=m_n^2\varphi_{n}(z),
\end{equation}
with
\begin{equation}
\label{Adagger}
\mathcal{A}=\frac{d}{dz}+\frac{\dot{\theta}}{\theta},\quad
\mathcal{A}^\dagger=-\frac{d}{dz}+\frac{\dot{\theta}}{\theta}.
\end{equation}
This equation assures the positive semi-definite of $m_n^2$ and equivalently, assures the stability of the solution.
Obviously, the zero mode ($m_0^2=0$) takes the form $\varphi_0\propto \theta^{-1}$. As we have pointed out in Ref.~\cite{ZhongLiu2013}, the scalar zero mode is always unlocalizable, no matter what value $\alpha$ takes. This can also be concluded from the shape of $U_S$ (see Fig.~\ref{figureScalarPertPotential}), from which we know there is no bound or resonant state in the spectrum of scalar graviton. So, same to the standard model, our model is free of the long range scalar fifth-force problem\footnote{This is because a localizable scalar zero mode corresponds to a new long range force gauge boson, which transmits a new force we have never seen before.}.

\section{Summary}
\label{section5}

We set an example for solving $K$-brane system via the first-order formalism. Our basic assumption is that the noncanonical kinetic term affect only the geometry of the space-time, while leave the scalar configuration unaffected. A novel first-order formalism is established to solve the type I model of Ref.~\cite{BazeiaGomesLosanoMenezes2009}.
An exact analytical solution was obtained by taking Sine-Gordon superpotential. We also studied the stability of the solution against linear field perturbations. The localization of fermion and graviton was analyzed.

Our study indicates that the stability conditions demand lower bound for the deviation parameter $\alpha$. On the other hand, the requirement of localizing bulk fermions imposes upper bound for the deviation parameter. Because for a given Yukawa coupling, a large deviation parameter can violate the localization condition of fermion zero mode.
In addition, we studied the localization of tensor and scalar gravitational perturbations. We found that the tensor zero mode is always localizable provided that the stability conditions are satisfied, while the scalar zero mode is always nonlocalizable. There is no sign for gravitational resonance either in the tensor or the scalar section.

Hopefully, the procedure we used here can be applied to other $K$-field models, such as cosmology, topological defects in various dimensions. Besides, the superpotential method also reminds us the possibility of supersymmetric extensions of the model. We leave all these topics to the future works.
\section*{Acknowledgments}

This work was supported by the Program for New Century Excellent Talents in University, the National Natural Science Foundation of China (Grants No. 11075065 and No. 11375075), and the Fundamental Research Funds for the Central Universities (Grant No. lzujbky-2013-18). Yuan Zhong was supported by the Scholarship Award for Excellent Doctoral Student granted by Ministry of Education, and the scholarship granted by the Chinese Scholarship Council (CSC). Zhen-Hua Zhao was upported by the National Natural Science Foundation of China (Grant No. 11305095) and the Natural Science Foundation of Shandong Province, China (Grant No. 2013ZRB01890), and Scientific Research Foundation of Shandong University of Science and Technology for Recruited Talents (Grant No. 2013RCJJ026).

%





\end{document}